\documentstyle[aps,multicol]{revtex}
\input epsf.sty
\begin{document} 
\draft

\title{Evidence for Two--Band Superconductivity from Break Junction 
Tunneling on MgB$_{2}$} 

\author{H.\ Schmidt,$^{1,2}$ J.\ F.\ Zasadzinski,$^{1,2}$ 
K.\ E.\ Gray,$^{1}$ D.\ G.\ Hinks$^{1}$} 

\address{$^{1}$Materials 
Science Division, Argonne National Laboratory, Argonne, IL 60439\\
$^{2}$Physics Division, Illinois Institute of Technology, Chicago, IL 60616} 

\date{\today} \maketitle

\begin{abstract}
Superconductor--insulator--superconductor tunnel junctions have been 
fabricated on MgB$_{2}$ that display Josephson and 
quasiparticle currents. These junctions exhibit a gap magnitude, 
$\Delta\sim2.5$ meV, that is considerably smaller than the BCS value, but 
which clearly and reproducibly closes near the bulk $T_{c}$. In 
conjunction with fits of the conductance spectra, these results are interpreted 
as direct evidence of two--band superconductivity.
\end{abstract}
\pacs{PACS numbers: 73.40.Gk, 74.50.+r, 74.70.Ad, 74.80.Fp}

\begin{multicols}{2}

The discovery of superconductivity in MgB$_{2}$ has led to intense 
research activity, but the nature of the energy gap, $\Delta$, has 
been elusive. 
Tunneling spectroscopy, which is the most direct measure of this 
quantity, has revealed a large spread of $\Delta$ values and 
considerable variation in its spectral shape. Sharp, BCS--like 
tunneling spectra have been observed in scanning tunneling microscopy 
(STM) with a surprisingly small $\Delta=2.0$ meV \cite{Rubio-Bollinger}.
Other STM and point--contact studies revealed double--peaked spectra at low 
temperatures \cite{Giubileo-PRL,Szabo} that were interpreted as evidence for 
two--gap superconductivity.  A provocative suggestion is that multiple 
gaps are a consequence of the coupling of distinct electronic bands \cite{Liu}. 
Our ability to fabricate 
superconductor--insulator--superconductor (SIS) break 
junctions has led to unique observations and we have gone beyond these initial 
reports to present more compelling 
evidence that MgB$_{2}$ is one of the rare 
examples of two--band 
superconductivity.  In addition, our identification of a weak higher--bias 
spectral feature has provided important insight into the nature of the inter--band 
coupling.

The simplicity of the crystalline structure in MgB$_{2}$ allows 
for {\em ab--initio} calculations of its electronic properties \cite{Kortus}, 
from which it is known that the Fermi surface 
consists of four sheets, two being two dimensional (2d) bonding $\sigma$--bands
and two being three dimensional (3d) bonding and antibonding $\pi$--bands. 
An and Pickett \cite{An} propose superconductivity to be 
driven by the 2d $\sigma$--bands, where electrons are strongly coupled 
primarily to the $E_{2g}$ phonon mode. This raises important questions of 
how superconductivity would manifest itself on the 3d sheets 
and how the tunneling density of states (DOS) would depend on the 
crystallographic orientation.

A more recent work \cite{Liu} treated the problem by reducing it to two 
distinct bands which, in the clean limit, leads to the appearance 
of two isotropic gaps, $\Delta_{2}\sim7.2$ meV and $\Delta_{1}\sim 2.4$ 
meV, associated with the 2d and 3d bands respectively.
The small gap, $\Delta_{1}$, on the 3d sheets is enhanced above its 
intrinsic value due to virtual phonon exchange (pair transfer) with the 
2d sheets and should persist up to the bulk $T_{c}$.  The results of our 
tunneling study address these issues in the following ways.    
First, the small gap feature is {\em unambiguously} tracked 
to high temperatures where it is still visible in the {\em raw data}, a 
key observation supporting two--band superconductivity.  
These junctions only probe the band with the small, induced gap suggesting the 
SIS configuration strongly favors tunneling between the 3d sheets. Second, 
a subtle spectral feature is observed in the conductance near 9 meV that resembles 
strong--coupling effects.  Using a theoretical, two--band model \cite{McMillan} that treats 
pair transfer and the quasiparticle self--energy on an equal footing, we have 
quantitatively fit this feature.  This indicates that self--energy effects 
originating from quasiparticle scattering between bands are important. 
Finally and importantly, using unique features 
of SIS junctions, we have more rigorously ruled out proximity effects which otherwise can mimic 
the temperature dependencies and the quasiparticle 
self--energy effects of two--band superconductivity.  

Compact samples of MgB$_{2}$ were formed from amorphous B powder 
(4N's purity) and high purity Mg.  The B powder was pressed into 
pellets under 6 kbar pressure.  These free standing pellets were reacted 
with Mg vapor at 850$^{\circ}$C for 2 hr in a BN container under 50 bar 
of Ar.  During the diffusion reaction the pellets broke up into irregularly 
shaped pieces several mm on a side.  The material typically showed 
$T_{c}=39$ K.  To obtain a clean and smooth surface, the samples were 
polished until a shiny surface was exposed.  No solvent was 
used and the samples were only cleaned in a flow of dry N$_{2}$ gas.

\begin{figure}
\centerline{
\begin{minipage}{\linewidth}
\centerline{\epsfxsize=0.9\linewidth 
\epsfbox{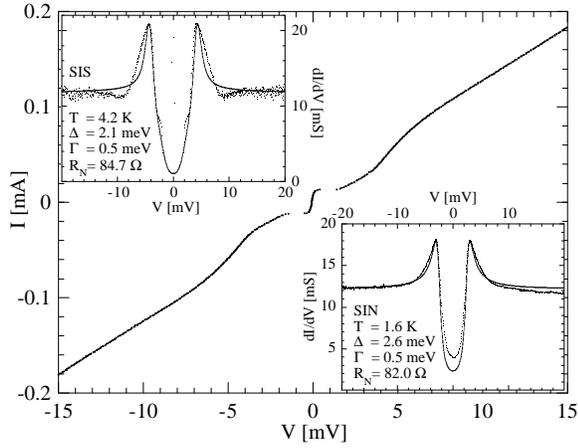}}
\vskip.15in \caption{Josephson tunnel junction on MgB$_{2}$ displaying 
an energy gap of 2.1 meV. Upper inset: $s$--wave fit to the 
conductance characteristic. Lower inset: Comparison to an SIN junction on 
sample B showing a comparable gap magnitude at 1.6 K.}
\label{jos}
\end{minipage}}
\end{figure}

The tunneling measurements were performed on two different samples (A and B) 
using a point--contact 
apparatus \cite{Ozyuzer} with a gold tip.  This technique yielded 
superconductor--insulator--normal metal (SIN) junctions as well as SIS break 
junctions, the latter being confirmed by fits to the temperature dependent 
conductance as well as the observation of Josephson currents.  Fig.\ \ref{jos} 
gives an example of a Josephson tunnel junction obtained on sample A. 
In the  current voltage  characteristic there is a 
distinct jump visible between a well--pronounced, slightly hysteretic  Josephson 
current and the quasiparticle branch which displays a gap 
feature at $eV =2\Delta$. In the $\frac{dI}{dV}$ \emph{vs.} 
$V$ spectra, sharp peaks 
are seen which can be  fit to a BCS $s$--wave model with $\Delta=2.1$ meV and 
$\Gamma=0.5$ meV. The fit is very good except for a weak spectral feature 
in the data near 9 mV, which we discuss
\noindent  later. The simultaneous observation 
of a well developed gap feature and a significant Josephson current  
establishes these junctions to be of SIS geometry. The gap value is 
significantly less than the BCS expectation of 
 5.9 meV, but consistent with the $I_{c}R_{N}$ product
of 2.2 meV for this junction (other junctions 
gave comparable values). Note that similarly low 
values of $I_{c}R_{N}$ were found in a recent study 
of Josephson junctions in MgB$_{2}$ \cite{Gonnelli} prepared by a 
completely different break junction method.  Although the focus of this 
study is on the SIS junctions, similar gap 
values were consistently found in SIN junctions (lower inset of Fig.\ 
\ref{jos}). Fits to a smeared BCS model 
yield $\Delta=2.6$ meV and $\Gamma=0.5$ meV, that are consistent with the 
parameters obtained for SIS junctions.

\begin{figure}
\centerline{
\begin{minipage}{\linewidth}
\vskip.07in
\centerline{\epsfxsize=0.9\linewidth 
\epsfbox{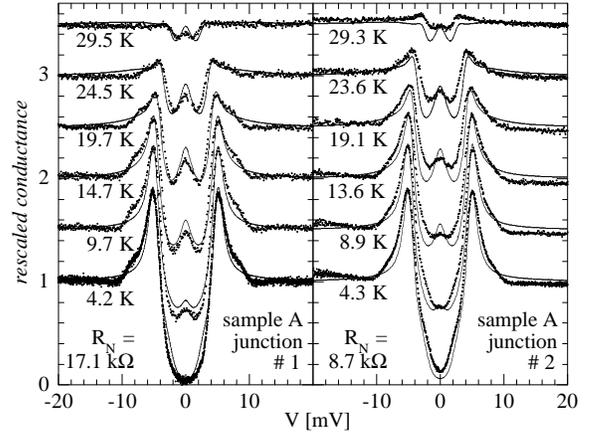}}
\vskip0.12in\caption{Temperature dependence of two high--resistance SIS
tunnel junctions together with $s$--wave fits reproducing all major 
features of the conductance characteristics, including the evolution 
of a zero--bias peak due to thermally activated current. Each spectrum 
is rescaled by the indicated $R_{N}$.}
\label{sis}
\end{minipage}}
\end{figure}

\begin{figure}
\centerline{
\begin{minipage}{\linewidth}
{\epsfxsize=0.95\linewidth 
\epsfbox{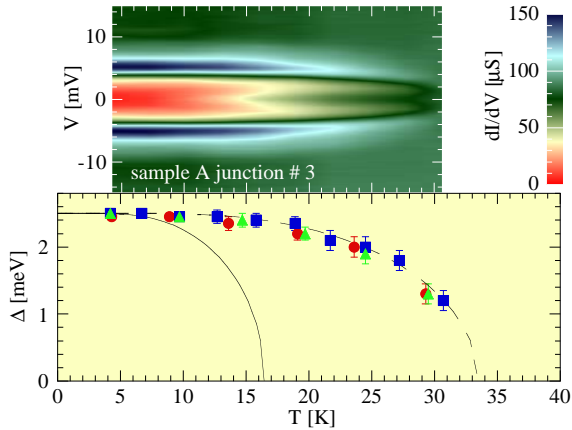}}
\vskip0.15in\caption{Top panel: Color map of SIS junction conductance 
spectra as a function of temperature.  Bottom panel: Temperature dependence 
of the small superconducting 
gap $\Delta_{1}$. Symbols are data extracted from SIS fits to the 
conductance spectra (different symbols represent three different 
experiments). The solid line gives $\Delta(T)$ according to BCS 
theory, using the expected ratio of $2\Delta(0)/k_{B}T_{c}$. The 
dashed line gives a {\em rescaled} BCS dependence with both $\Delta$ 
and $T_{c}$ adjusted to fit the experimental data.}
\label{deltaT}
\end{minipage}}
\end{figure}

Many SIS junctions were obtained at different locations on samples A 
and B.  With increasing resistance of the junction the 
Josephson currents decrease, and for contact resistances higher 
than $\sim10$ k$\Omega$ no supercurrent is visible at zero bias. 
Nevertheless, the junctions could be easily identified as SIS by the 
evolution of the conductance spectra with increasing temperature.  
SIN junctions show a rapid decrease in peak height at $eV=\Delta$ with 
increasing temperature, while  the peak position remains approximately constant. 
In contrast, the gap structure of SIS junctions remains sharp even for elevated 
temperatures, and the $2\Delta$--peak position follows the  closing of 
the gap at $T_{c}$.
We traced three such junctions formed on different parts of the MgB$_{2}$ 
sample A up to temperatures of $\sim30$ K, and two of these sets of data 
are shown in Fig.\ \ref{sis}, while the third is shown as a color map in 
Fig.\ \ref{deltaT}.  What is immediately evident in the raw data is that the 
small gap persists up to 30 K.  Each curve shown in Fig.\ \ref{sis} was 
divided by $R_{N}$ (the high bias resistance of the 4.2 K data) and then fit to an 
$s$--wave gap SIS model using the measured temperature.  The smearing 
parameter $\Gamma$ was adjusted once to reproduce the low--temperature 
data and then held fixed. This was done to prove 
the
\noindent  ability of the model to account for all major features of the temperature 
dependent data using $\Delta$ as the only adjustable parameter. The fits shown in Fig.\ \ref{sis} reproduce the peak heights, the shape of 
the in--gap conductance as well as the evolution of the zero--bias peak 
that is due to 
thermally activated quasiparticle current.  This zero--bias feature cannot 
appear in SIN tunneling but is a known feature of SIS tunneling. The 
magnitude of this central peak 
is surprisingly high, and this is caused by the unusual case of a 
small gap which persists to comparatively high temperatures (much 
higher than the BCS $T_{c}$ connected with this gap) thus 
giving rise to unusually large thermal activation of quasiparticles. The 
BCS $T_{c}$ for a low temperature gap of $\sim2.5$ meV is  
below 17 K, and the gap seen here is still clearly 
developed at a temperature around 30 K. This rules out a lower 
$T_{c}$ on the sample surface as an explanation of the small gap value as 
this would display a second transition near 17 K, which is not 
observed (see Fig.\ \ref{deltaT}).  Instead, these features 
indicate a bulk property is being measured that clearly deviates from 
strong coupling or BCS weak coupling theory.

After taking the first set of temperature--dependent data (junction 
\#1) the point contact tip was twice mechanically retracted over 
$\sim100$ $\mu$m and two new junctions (\#2 and \#3) were formed on the same 
sample (A), showing the identical spectral shapes and temperature dependence. 
Since the thread mechanism of the tip  approach does not preserve the 
microscopic lateral 
position on the sample, these junctions have to be  
regarded as entirely independent and their consistency thus proves
the  reproducibility of these observations. In addition it will be shown 
that SIS junctions on sample B display nearly identical low--$T$ 
characteristics. Further tests, which included etching 
of the Au tip, suggested these junctions were formed by breaking off 
an MgB$_{2}$ crystal fragment which then forms an SIS junction with 
the bulk material. 
\noindent This procedure of creating SIS break junctions is well 
known for point--contact tunneling into BSCCO  \cite{DeWilde}. The ease with 
which reproducible SIS junctions were formed was quite surprising and this 
requires us to critically review our previously published data which showed 
similar spectral shapes at 4.2 K \cite{Schmidt}, but no $T$--dependence 
was measured. The data presented there were interpreted in terms of SIN 
tunneling with a gap value of $4.3-4.6$ meV and no smearing parameter $\Gamma$.  
This analysis is similar in shape to an SIS characteristic with 
$\Delta=2.5$ meV and $\Gamma=0.5$ meV (typical parameters used here to 
fit the data e.g.\ in Fig.\ \ref{sis}), leading to the possibility 
that some of the junctions in that work were in fact SIS type.  
This would bring the measured gaps of that work in line with what was 
consistently observed here.      

The values for $\Delta_{1}(T)$ extracted from the fits to the conductance data 
on these three junctions are shown in Fig.\ \ref{deltaT}.  The uncertainty 
in the measured gap values is given by the error bars and there is excellent 
reproducibility among the three junctions. The solid line 
gives the BCS dependence for a low temperature gap of 2.5 meV with the 
expected $T_{c}=\Delta_{1}(0)/(1.76k_{B})=16.5$ K.  
Comparison of the data with the BCS fit shows a clear anomaly, the small 
gap persists up to temperatures far beyond the expected $T_{c}$.  The dashed 
line gives a {\em rescaled} BCS $\Delta(T)$ with the $T_{c}$ adjusted to fit 
the experimental data. This leads to an extrapolated $T_{c}$ near the bulk 
value for MgB$_{2}$ but we want to strongly emphasize, that there
is no basis for this type of rescaling within BCS theory.  This should be 
contrasted with strong--coupling effects which commonly lead to a zero 
temperature gap that exceeds the BCS value resulting in an 
\emph{enhanced} gap ratio $2\Delta(0)/k_{B}T_{c}$.  Here, this ratio 
is much \emph{smaller} than the expected BCS value.   

We propose that the data of Fig.\ \ref{deltaT} together with our 
analysis of the conductance spectra presented below are compelling evidence of 
two--band superconductivity as suggested by Liu \emph{et al.} 
\cite{Liu}.  The absence of any evidence for a second transition in Fig.\ 
\ref{deltaT} indicates that $\Delta_{1}$ is primarily 
induced via coupling to the 2d band.  We believe that the persistent and clear 
observation of only $\Delta_{1}$ in our SIS and SIN junctions is due to the 
much higher probability of tunneling into the 3d band. This band 
accounts for 58\% of the total DOS \cite{Liu} which is insufficient to account 
for its preference. It rather is the dominance 
of tunneling by electrons with momenta normal to the barrier which 
favors the 3d band.  Assuming a random orientation, 
there is a relatively low probability of being properly aligned with the 2d band.  

The low temperature conductance spectra reveal a weak but reproducible structure
near 9 mV which we believe is related to the large gap, $\Delta_{2}$, but is not 
due to direct tunneling into the 2d band.  This conclusion is supported by a 
calculation of the quasiparticle DOS on each sheet using the McMillan 
tunneling model \cite{McMillan}. This model simulates a coupled, two--band 
system by including the BCS--type, virtual phonon coupling (pair exchange) 
between bands \cite{Suhl,Noce} and also self--energy effects from interband 
quasiparticle exchange. The model requires the solution of two 
simultaneous equations:
\begin{equation}
\Delta_{1}(E)={\Delta_{1}^{\rm ph}+
{\Gamma_{1}\Delta_{2}(E)/
\sqrt{\Delta_{2}^{2}(E)-(E-i\Gamma^{*}_{2})^{2}}}\over
1+{\Gamma_{1}/
\sqrt{\Delta_{2}^{2}(E)-(E-i\Gamma^{*}_{2})^{2}}}}
\end{equation}
for the energy dependence of the two gaps, where the second equation 
is obtained by interchanging the subscripts 1 and 2. These functions 
$\Delta_{1,2}(E)$ are subsequently used to compute the DOS 
in both bands via the usual BCS expression. Convolution of these DOS then 
yields the desired SIS conductance characteristics. The model includes six 
parameters: the intrinsic pairing gaps on both bands in the absence of any 
interband coupling, $\Delta_{1,2}^{\rm ph}$,
two scattering rates, $\Gamma_{1,2}$, related inversely to the times spent 
in each band prior to scattering to the other, and two smearing parameters,
$\Gamma^{*}_{1,2}$, which were added to account for 
lifetime effects within each band.
Rather than treat each parameter as free, we fix the intrinsic pairing 
gaps by considering first principle calculations for MgB$_{2}$ 
\cite{Liu}, {\em viz.}\ $\Delta_{1}^{\rm ph}=0$ and $\Delta_{2}^{\rm ph}=7.2$ meV 
\cite{deltas}. The first assumption is further justified by noting 
that only a single transition is observed in Fig.\ \ref{deltaT}. The 
observed induced gap magnitude is then adjusted by appropriate choice 
of $\Gamma_{1,2}$. These parameters are highly interdependent and 
cannot separately be determined. Good fits can be obtained for ratios 
$\Gamma_{2}/\Gamma_{1}$ between zero and $\sim0.5$. We choose 0.25 as 
representative \cite{gammaratio}.
The smearing parameters finally are needed to account for 
the broadening and the in--gap current. 

\begin{figure}
\centerline{
\begin{minipage}{\linewidth}
\centerline{\epsfxsize=0.9\linewidth
\epsfbox{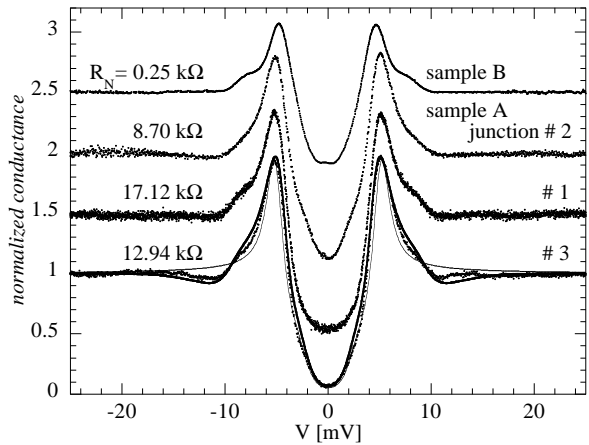}}
\vskip0.15in\caption{Low temperature SIS tunneling spectra. The measured data 
(dots) is normalized to a smooth background. For one set of data the fit 
(thin solid line) to a small (intrinsic) 
BCS $s$--wave gap is shown along with a fit (thick solid line) to an induced 
gap. The parameters used are $\Delta_{1}^{\rm ph}=0$â,
$\Delta_{2}^{\rm ph}=7.2$ meV, $\Gamma_{1}=4.0$ meV, $\Gamma_{2}=1.0$ meV,
$\Gamma_{1}^{*}=0.5$ meV and $\Gamma_{2}^{*}=1.0$ meV. See text for 
details.}
\label{mcm}
\end{minipage}}
\end{figure}

The low--temperature spectra from different junctions on samples A and B 
are shown in Fig.\ \ref{mcm} to demonstrate  the reproducibility of the data. 
Note that all
\noindent  characteristic 
features of the data taken on sample B, 
such as the gap magnitude and the additional structure outside the 
conductance peaks, 
coincide with those from sample A. For one junction (\# 3), the 
result of a two--band fit is  given along with the BCS model including 
a smearing  of $\Gamma=0.5$ meV. 
The BCS fit convincingly reproduces the in--gap conductance and the quasiparticle 
peak, but it cannot account for the features near 9 mV, which are 
consistently seen in these SIS junctions. The two--band fit is 
close to the BCS behavior for low bias, but it also reproduces the 
shoulder and dip at the observed energy in agreement with the data.  
The higher energy spectral feature is an effect of the large gap, 
$\Delta_{2}$, on the quasiparticle self--energy in the 3d band DOS.  Note 
that the dip drops below unity, a characteristic feature of quasiparticle 
self--energy effects, similar to phonon structures, but which cannot be 
achieved by arbitrarily adding two BCS DOS from the two bands.  The 
subtlety of the feature near 9 meV, as well as the dip, is more easily 
understood as an intrinsic feature of the 3d DOS.  Our data show no direct 
contribution from the 2d band \cite{2dbands}.

The same formalism used above to calculate the conductances is valid for a 
proximity sandwich, P, consisting of N and S layers that are coupled by 
a tunnel barrier I \cite{McMillan}, {\em i.e.} P $\equiv$ NIS. 
Well--coupled N and S layers 
result in spectral features inconsistent with our observations 
\cite{Arnold}. For P $\equiv$ NIS, the small gap in the N region has 
been shown to decrease exponentially with barrier thickness \cite{Gray}.  
To explain the small gap of MgB$_{2}$ as a surface 
proximity effect ({\em i.e.}, PIP' for our SIS geometry) would produce 
a noticeable splitting of the zero--bias peak (Fig.\ \ref{sis}) unless the 
gaps in P and P' differed by less than $\sim20$\%.  This is highly unlikely 
as it would require \cite{Gray} the tunneling barrier thicknesses 
in P for all MgB$_{2}$ samples to be within $\sim0.5$ \AA.  Thus it is our SIS 
geometry that allows us to rule out proximity effects.

In conclusion,  we have reproducibly observed a small energy gap, 
$\Delta_{1}\sim2.5$ meV, which smoothly closes near the bulk $T_{c}$.  This 
observation, along with a detailed analysis of the conductance spectra, is
indicative that MgB$_{2}$ is an example of the rarely observed 
phenomenon of two--band superconductivity.
While the measured gap values are 
consistent with the first principles two--band model that assumes pair 
exchange between bands \cite{Liu}, our spectra are providing strong 
evidence that interband quasiparticle exchange is important.  
The data are inconsistent with surface layers of reduced $T_{c}$ or with 
proximity effects.   

We gratefully acknowledge helpful discussions with M.\ R.\ Norman, 
G.\ B.\ Arnold, B.\ Jank$\rm\acute{o}$ and C.\ P.\ Moca. 
This research is supported by the U.S.\  Department of Energy,
Basic Energy Sciences---Materials Sciences, under contract \# 
W--31--109--ENG--38.

\newpage
\end{multicols}

\begin{references}
\bibitem{Rubio-Bollinger} G.\ Rubio--Bollinger, H.\ Suderow and S.\ 
Vieira, Phys.\ Rev.\ Lett.\ \textbf{86}, 5582 (2001).
\bibitem{Giubileo-PRL} F.\ Giubileo {\em et al.}, Phys.\ Rev.\ 
Lett.\ \textbf{87}, 177008 (2001).
\bibitem{Szabo} P.\ Szab$\rm\acute{o}$ {\em et al.}, Phys.\ Rev.\ 
Lett.\ \textbf{87}, 137005 (2001).
\bibitem{Liu} A.\ Y.\ Liu, I.\ I.\ Mazin and J.\ Kortus, Phys.\ Rev.\ Lett.\ 
\textbf{87}, 087005 (2001).
\bibitem{McMillan} W.\ L.\ McMillan, Phys. Rev. \textbf{175}, 537 (1968).
\bibitem{Kortus} J.\ Kortus {\em et al.}, Phys.\ Rev.\ Lett.\ 
\textbf{86}, 4656 (2001). 
\bibitem{An} J.\ M.\ An and W.\ E.\ Pickett, Phys.\ Rev.\ Lett.\ 
\textbf{86}, 4366 (2001).
\bibitem{Ozyuzer} L.\ Ozyuzer, J.\ F.\ Zasadzinski and K.\ E.\ Gray, 
Cryogenics \textbf{38}, 911 (1998).
\bibitem{Gonnelli} R.\ S.\ Gonnelli {\em et al.}, 
Phys.\ Rev.\ Lett.\ \textbf{87}, 097001 (2001).
\bibitem{DeWilde} Y.\ DeWilde {\em et al.},
Phys.\ Rev.\ Lett.\ \textbf{80}, 153 (1998).
\bibitem{Schmidt} H.\ Schmidt {\em et al.}, 
Phys.\ Rev.\ B \textbf{63}, 220504(R) (2001). 
\bibitem{Suhl} H.\ Suhl, B.\ T.\ Matthias and L.\ R.\ Walker, 
Phys.\ Rev.\ Lett.\ \textbf{3}, 552 (1959).
\bibitem{Noce} C.\ Noce and L.\ Maritato, 
Phys.\ Rev.\ B \textbf{40}, 734 (1989).
\bibitem{deltas} The calculated coupling constant $\lambda_{B}=0.45$ 
\cite{Liu} for the 3d--band yields an intrinsic pairing with $T_{c}<4.2$ K. 
It therefore is safe to neglect $\Delta_{1}^{\rm ph}$ for our purpose.
\bibitem{gammaratio} Since $\Gamma_{1,2}$ are interband scattering rates, their ratio is 
expected to be related to the ratio of the DOS in both bands, $N_{1}/N_{2}$. 
However, due to the complicated geometry of the Fermi surface, it is not 
evident to us that all quasiparticles participate in the interband scattering 
and we therefore do not expect to find $\Gamma_{2}/\Gamma_{1}=N_{1}/N_{2}$.
\bibitem{2dbands} In principle, three channels contribute in parallel to the 
total current in two--band SIS junctions. In the order of increasing tunneling 
probability, these are 2d--2d, 2d--3d and 3d--3d, corresponding to
quasiparticle peaks at $eV=2\Delta_{2}$, $\Delta_{1}+\Delta_{2}$ and 
$2\Delta_{1}$. Only the peaks of the dominant 3d--3d spectrum are observed.
\bibitem{Arnold} G.\ B.\ Arnold, Phys.\ Rev.\ B \textbf{18}, 1076 
(1978).
\bibitem{Gray} K.\ E.\ Gray, Phys.\ Rev.\ Lett.\ \textbf{28}, 959 (1972).  
\end{references}
\end{document}